\documentclass[aps,pra,twocolumn]{revtex4-1}

\usepackage{graphicx}
\usepackage{xcolor}
\usepackage{soul}

\begin{document}

\title{The optimal reordering of measurements for photonic quantum tomography}
\author{Radim Ho\v{s}\'ak}
\affiliation{Department of Optics, Palack\'{y} University, 17. listopadu 12, 77146 Olomouc, Czech Republic}
\author{Robert St\'arek}
\affiliation{Department of Optics, Palack\'{y} University, 17. listopadu 12, 77146 Olomouc, Czech Republic}
\author{Miroslav Je\v{z}ek}
\email{jezek@optics.upol.cz}
\affiliation{Department of Optics, Palack\'{y} University, 17. listopadu 12, 77146 Olomouc, Czech Republic}

\begin{abstract}
Quantum tomography is an essential method of the photonic technology toolbox and is routinely used for evaluation of experimentally prepared states of light and characterization of devices transforming such states. The tomography procedure consists of many different sequentially performed measurements. We present considerable tomography speedup by optimally arranging the individual constituent measurements, which is equivalent to solving an instance of the traveling salesman problem. As an example, we obtain solutions for photonic systems of up to five qubits, and conclude that already for systems of three or more qubits, the total duration of the tomography procedure can be halved. The reported speedup has been verified experimentally for quantum state tomography and also for full quantum process characterization up to six qubits, without resorting to any complexity reduction or simplification of the system of interest. Our approach is versatile, and reduces the time of an input-output characterization of optical devices and various scattering processes as well.
\end{abstract}


\maketitle

\section{Introduction}

\begin{figure}[t!]
\centerline{\includegraphics[width=0.94\columnwidth]{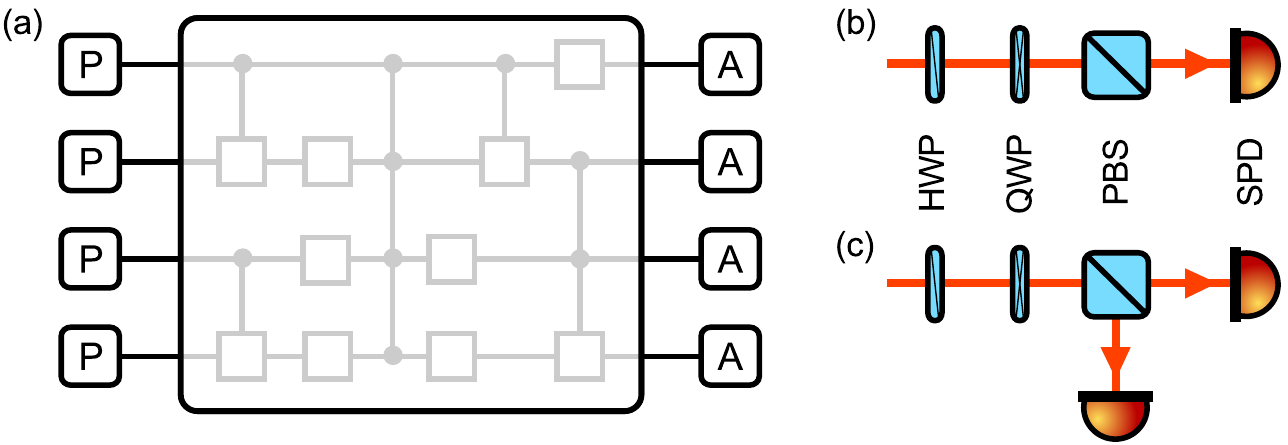}}
\caption{(a) Complete characterization of a quantum circuit based on probing (P)
the input qubits by all possible combinations of quantum states selected
from a quorum and performing full analysis (A) of the output qubits.
The analysis of the output qubit consists of projections of the
qubit state onto quorum states. (b) For photonic circuits the analysis (A) is
often polarization-encoded and formed by a sequence of wave plates
(half-wave, HWP; quarter-wave, QWP), a polarization beam splitter (PBS),
and a single-photon detector (SPD). The wave plates are rotated to perform
the projections to particular polarization states. (c) When both outputs
of the PBS are detected by the SPDs, two orthogonal projections are measured
at the same time, reducing the duration of the measurement at the expense
of a higher number of SPDs employed. The preparation stage (P)
is constructed in a similar way using a proper single-photon source,
a polarizer, and a sequence of wave plates.}
\label{fig:polar_tomo}
\end{figure}

Quantum photonic technologies have the potential to revolutionize
information and communication systems, employing non-classical
states of light to encode, manipulate, and transmit information \cite{OBrien2009rev,Sciarrino2018}.
There has been a significant progress in secure communications,
photonic simulations, and other recently emerging fields, especially
with the advent of novel sources of entangled states \cite{Trotta2018,Predojevic2018} and quantum integrated circuits \cite{Sciarrino2017,O'Brien2018,Walmsley2018}.
The complexity of the quantum states, and their transformations, grows exponentially
with the number of quantum bits (qubits).
Quantum information processing exploits this scaling to realize quantum algorithms and circuits, showing
a potential advantage over their classical counterparts.
In order to evaluate the performance of the quantum state sources and the quantum circuits,
it is necessary to perform their characterization.
Quantum state tomography
\cite{VogelRisken1989,Jones1994,Hradil1997,Banaszek1999,Munro2001},
developed to estimate quantum state based on measurement outputs,
is an essential part of the quantum information processing toolbox
and is routinely used for characterization of experimentally prepared
quantum states of light \cite{ParisRehacek2004,Lvovsky2009}.
Also, quantum process tomography \cite{Zoller1997,ChuangNielsen1997,Fiurasek2001}
is a crucial tool for full characterization of a device that transforms quantum states,
like an on-chip interferometric network or a quantum circuit
processing qubits in a quantum register, see Fig.~\ref{fig:polar_tomo}(a).
Despite being a much more complicated
procedure, it was shown that quantum process tomography is equivalent
to quantum state tomography performed on a larger parameter space
\cite{DAriano2001,White2003,Hradil2003}.

A photonic dual-rail qubit encodes information in two modes, such as horizontal
and vertical polarization or lower and upper path of an interferometer,
which serve as computational basis states \cite{Kok2007,Sciarrino2018}.
Estimating an unknown state of a qubit system by means of quantum state tomography
requires projecting the state onto quorum states, often chosen as
tensor products of single-qubit Pauli eigenstates \cite{Altepeter2005},
a minimum set of projections \cite{Rehacek2004},
or randomly selected \cite{Eisert2010,White2011,Weinfurter2014,Eisert2017}.
Despite the choice of set of tomographic measurements,
one has to sequentially adjust the individual projections using programmable
analyzers, often realized using birefringent wave plates in rotation mounts,
voltage controlled liquid crystal modulators, or various phase shifters,
see Figs.~\ref{fig:polar_tomo}(b) and~\ref{fig:polar_tomo}(c).
The analyzers need to be readjusted when switching from one measurement to another,
which can be time-consuming. It is also generally the case that the transition time
depends on the particular measurement sequence. Consequently, it is reasonable to ask
whether it is possible to optimize the order of tomographic
measurements with respect to the total time spent on readjusting the
analyzers \cite{Altepeter2005}. We are then faced with an
instance of the traveling salesman problem (TSP) \cite{Lawler1985}.
For quantum process tomography, the optimization can be performed at the measurement
stage as well as the preparation stage.
We will then further discuss the optimization of the measurement sequence, knowing that the results also hold for state preparation.
Indeed, the preparation sequence can always
be the subject of optimization even when the measurement is implemented without the
necessity of analyzer readjusting \cite{Kurtsiefer2006,Szameit2018}. The reduction of
quantum tomography duration is highly desirable as it also reduces the overall
setup drift and other systematic errors, which typically yields improved
quality of the quantum operation or state characterized \cite{Langford2013,BlumeKohout2013}.

In this work, we solve the measurement order optimization for the case of
polarization tomography with wave plates in rotation mounts. The optimization for
other physical realizations of analyzers or preparation stages can be performed
analogously. We illustrate our approach on a one-qubit system, and then find
the optimal strategy for systems of increasing numbers of qubits, going up to five.
We quantify the time saved on reorienting the wave plates in terms of the speedup
factor and find an increasing trend with each added qubit. We also experimentally
demonstrate the feasibility of the speedup in systems of up to six qubits,
where we achieve a speedup close to the predicted one.
Furthermore, we discuss possible optimization for path-encoded qubits on an optical chip and random-measurement tomographic techniques.

\section{One-qubit case}

We start with a simple example of one-qubit state tomography. We consider a
tomography implementation with a half-wave plate, a quarter-wave plate, and a
polarizing beam splitter with a detector in one of its output ports (further
called the six-state scheme) shown in Fig.~\ref{fig:polar_tomo}(b), where
tomography is realized by projecting on six polarization states in three
mutually unbiased bases. These polarization states are namely the horizontal
(H), vertical (V), diagonal (D), anti-diagonal (A), right-handed circular (R),
and left-handed circular (L). The wave plates are in motorized rotation mounts,
and projections on each of these states can be measured after orienting the
wave plates so that their axes of birefringence are at specific angles with the
plane of horizontal polarization of the measured light. Conventionally, the
projections are measured in the order in which the polarization states H
through L were presented here unless a Gray-code-like ordering is used
where just one wave plate is allowed to move during each readjustment
\cite{Altepeter2005}. The wave plate angles for all the projections are
shown in Table \ref{tab:tomo_angles}.
During transitions between individual measurements, it is necessary to rotate the
wave plates by different angles, which in turn takes different amounts of time
for the motorized mounts to realize. As more than one wave plate will generally
have to be rotated, we compare all of the individual wave plate rotation times
and take the maximum as the transition time.

Also note that while it is the total time spent on transitions that we
ultimately seek to minimize, we will continue to specify the problem in terms
of the wave plate rotation angles (absolute valued), and, where convenient,
still refer to time instead of angle. This is so that the problem
specification and solution are not dependent on the particular rotation mounts used. 
The two formulations, temporal and angular, are indeed equivalent if the rotation
time of the mount scales linearly with the angle traveled, which is practicaly
the case. The temporal formulation has to be used for different polarization
analyzers employed, i.e. liquid crystal modules or piezo-based fiber polarization
controllers, where the angle is not defined and the transition time does not
scale linearly with control voltages.

\begin{table}[t!]
	\centering
	\caption{One-qubit polarization tomography (the six-state scheme). The H, V,
    D, A, R, and L projections are shown with their corresponding wave plate
    angles in degrees.}
    \medskip
    \begin{tabular}{c | c c}
        & $\alpha_{\textrm{\scriptsize\,HWP}}$ & $\alpha_{\textrm{\scriptsize\,QWP}}$ \\ 			\hline
      H & 0 & 0 \\
      V & 45 & 0 \\
      D & 22.5 & 0 \\
      A & -22.5 & 0 \\
      R & 0 & 45 \\
      L & 0 & -45
    \end{tabular}
    \label{tab:tomo_angles}

\end{table}

\begin{table}[t!]
	\centering
    \caption{The TSP specification using the adjacency matrix consists of the maximal angles rotated by any wave plate during transitions
    between projections.}
    \medskip
    \begin{tabular}{c | c c c c c c}
        & H & V & D & A & R & L \\ \hline
      H & 0 & 45 & 22.5 & 22.5 & 45 & 45 \\
      V & 45 & 0 & 22.5 & 67.5 & 45 & 45 \\
      D & 22.5 & 22.5 & 0 & 45 & 45 & 45 \\
      A & 22.5 & 67.5 & 45 & 0 & 45 & 45 \\
      R & 45 & 45 & 45 & 45 & 0 & 90 \\
      L & 45 & 45 & 45 & 45 & 90 & 0
    \end{tabular}
    \label{tab:tomo_adj}
\end{table}

Now, with a set of states to project onto, and the corresponding wave plate angles,
we can proceed to the optimization. In our case, the number of projections to be
measured during tomography is $p=6$ and the number of all the possible
permutations of their order is $p! = 720$. It is then feasible to find the
measurement order of the least total transition time using a brute-force method.
This approach, however, does not scale well with the number of qubits in the
quantum system of interest. For example, in the case of a two-qubit system, the number of
measurements is $p^2=36$ and the number of their possible permutations is
$36!\approx 10^{41}$. Our only option then is to solve the TSP using the
state-of-the-art branch and bound algorithms \cite{Lawler1985}.

The TSP is commonly specified using the adjacency matrix.
Its element $c_{ij}$ is in our case given as the maximal
angle traveled by any wave plate during a transition from measurement $i$ to
measurement $j$. The diagonal elements are zero, and the matrix is symmetrical.
The adjacency matrix for the one-qubit, six-state scheme case is shown in
Table \ref{tab:tomo_adj}. This matrix can then be given as an input to any TSP
solver compliant with the standard defined by TSPLIB \cite{Reinelt1991}. We
made use of the Concorde solver \cite{Concorde}. The solver finds the
shortest \emph{cycle} of measurements, meaning that the optimization takes into
consideration also the transition from the final measurement back to the initial
one (the preparation for the next run of tomography). The resulting optimal measurement sequence is shown in Table \ref{tab:tomo:1qubit}.

\begin{table}
   \caption{The optimal projection measurement sequence for one-qubit tomography.}
   \begin{center}
   \begin{tabular}{c c c c c c}
   H & L & A & R & V & D
   \end{tabular}
   \end{center}
  \label{tab:tomo:1qubit}
\end{table}

To quantify the reduction in the total transition time $\tau_{\rm TSP}$ of
the TSP-optimized order of measurements compared to the conventional order,
where the time spent on transitions is $\tau_{\rm conv}$, we use the speedup
factor $s = \tau_{\rm conv} / \tau_{\rm TSP}.$
The time $\tau_{\rm conv}$ can be readily computed from the adjacency matrix as
$\tau_{\rm conv} = \left(\sum_{i = 1}^5 c_{i, i+1}\right) + c_{6,1},$
the last term being the transition from the final to the initial measurement,
completing the cycle.
For the one-qubit case, we found that the total angular duration of the cycle
of transitions with the conventional order of measurements was
292{.}5$^\circ$, whereas for the TSP-optimized order of measurements it was
225$^\circ$. The same result is obtained using the brute-force method.
The speedup factor for single-qubit tomography is therefore $s = 1.3$.
The same speedup can also be exploited in classical polarimetry with a strong
optical signal.

\section{More qubits}

For the tomography of a multi-qubit system it is not enough to perform the
one-qubit procedure for each individual qubit. It is necessary to measure the
projection onto every combination of the single-qubit polarization states.
This requirement leads to a $p^n$ scaling of the number of
measurements, $p$ being the number of measurements for a
single-qubit state, and $n$ being the number of qubits in the quantum system.
This naturally affects the duration of the entire tomography
measurement, and gives an even greater incentive to reduce it.

\begin{figure}[t!]
\centering{
  \includegraphics[width=0.96\columnwidth]{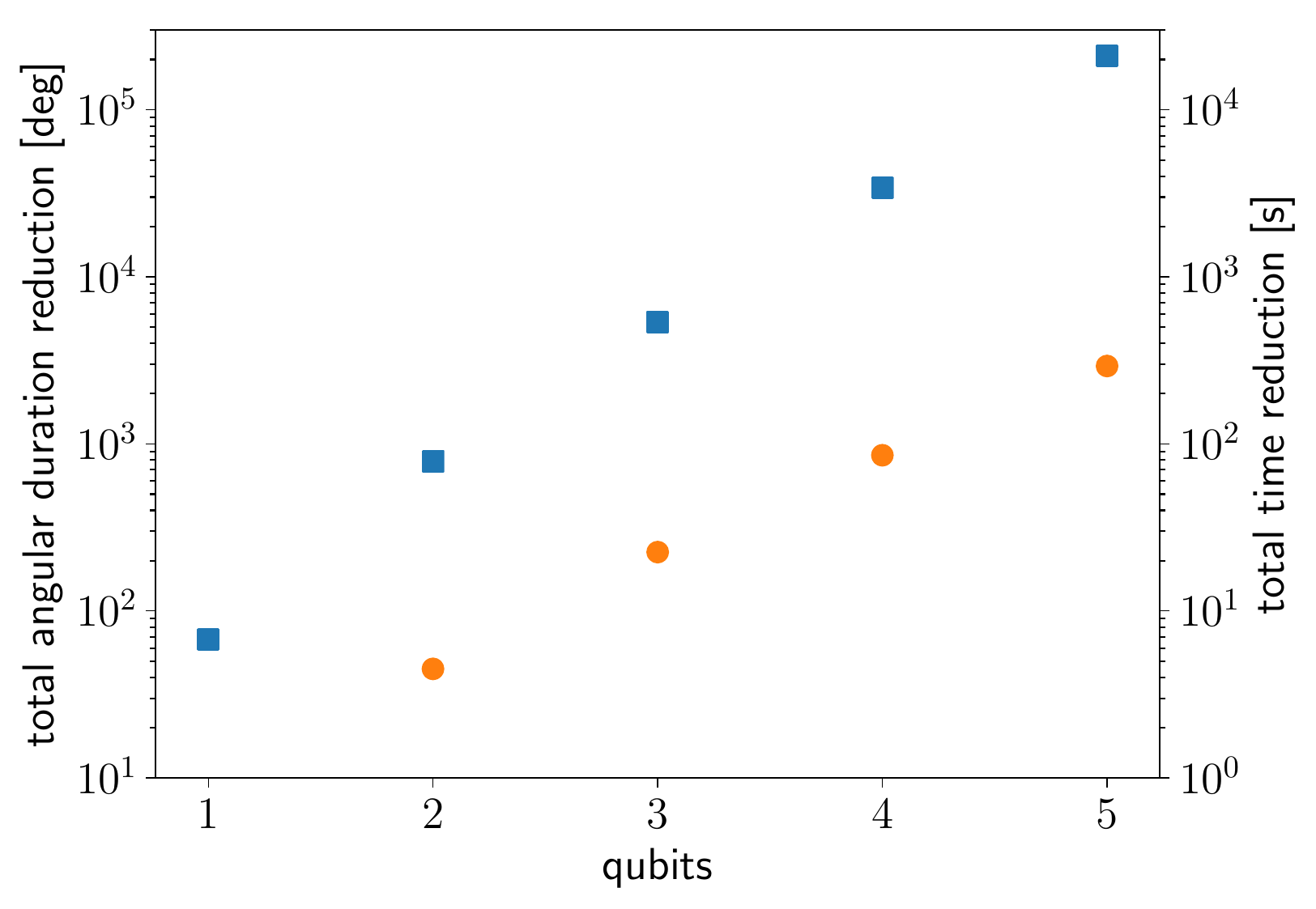}
}
\caption{The reduction of the total angular duration of tomography achieved when TSP-optimized sequence of measurements is used instead of the conventional one, for the six-state scheme (blue square) and the three-base scheme (orange circle). The temporal duration reduction, shown on the right-hand scale, assumes rotation mounts with 10 deg/s speed. The conventional one-qubit, three-base tomography is already optimal, and the corresponding data point of zero reduction is not shown in the plot.}
\label{fig:tau}
\end{figure}

As the number of possible permutations of the order of the tomographic
measurements is given by a factorial, we were forced to abandon the brute force
method and rely on a TSP solver, where we were able to obtain the optimal
measurement orders for up to five-qubit systems in a similar fashion as in the
previous single-qubit example. The optimal sequence for two-qubit tomography is shown in Table~\ref{tab:tomo:2qubit}. Measurement sequences for systems of higher numbers of qubits are too long to show in print. They are instead stored online \cite{Hosak2018}. The size of the adjacency matrix scales as $p^n
\times p^n$. For the six-state scheme and a five-qubit system, this means
that a $7776 \times 7776$ matrix was required for the specification of the TSP.
Matrices of such size were generated automatically by a computer program
\cite{Hosak2018}.

\begin{table}
   \caption{The TSP-optimized projection measurement sequence for two-qubit tomography, to be read left-to-right, top-to-bottom.}
   \begin{center}
   \begin{tabular}{c c c c c c c c c c c c}
   HH & HA & AH & AD & AV & AA & RR & RL & RD & RH & RA & RV \\
   AL & AR & HL & HR & DR & DL & VL & VR & LR & LL & LA & LV \\
   LD & LH & VV & VD & VH & VA & DD & DV & DA & DH & HV & HD \\
   \end{tabular}
   \end{center}
  \label{tab:tomo:2qubit}
\end{table}

Additionally, apart from the six-state scheme, another configuration has
also been considered. Using a detector in each of the output beam splitter ports
as shown in Fig.~\ref{fig:polar_tomo}(c), we can measure the projections onto
both of the orthogonal polarization states in a given basis simultaneously.
This allows for one-qubit tomography to consists of three, and not six,
measurements. Not the individual states, but the bases, are readjusted
leading to simultaneous measurements of the H and V, D and A, and R and L
states, with the wave plate angles the same as for the H, D, and R state,
respectively. We will further refer to this configuration as the three-base scheme.
It is important to note that employing the three-base scheme is limited to quantum
state tomography and cannot be utilized for probe state preparation in quantum
process tomography or any input-output system characterization.

\begin{figure}[t!]
\centering{
  \includegraphics[width=0.96\columnwidth]{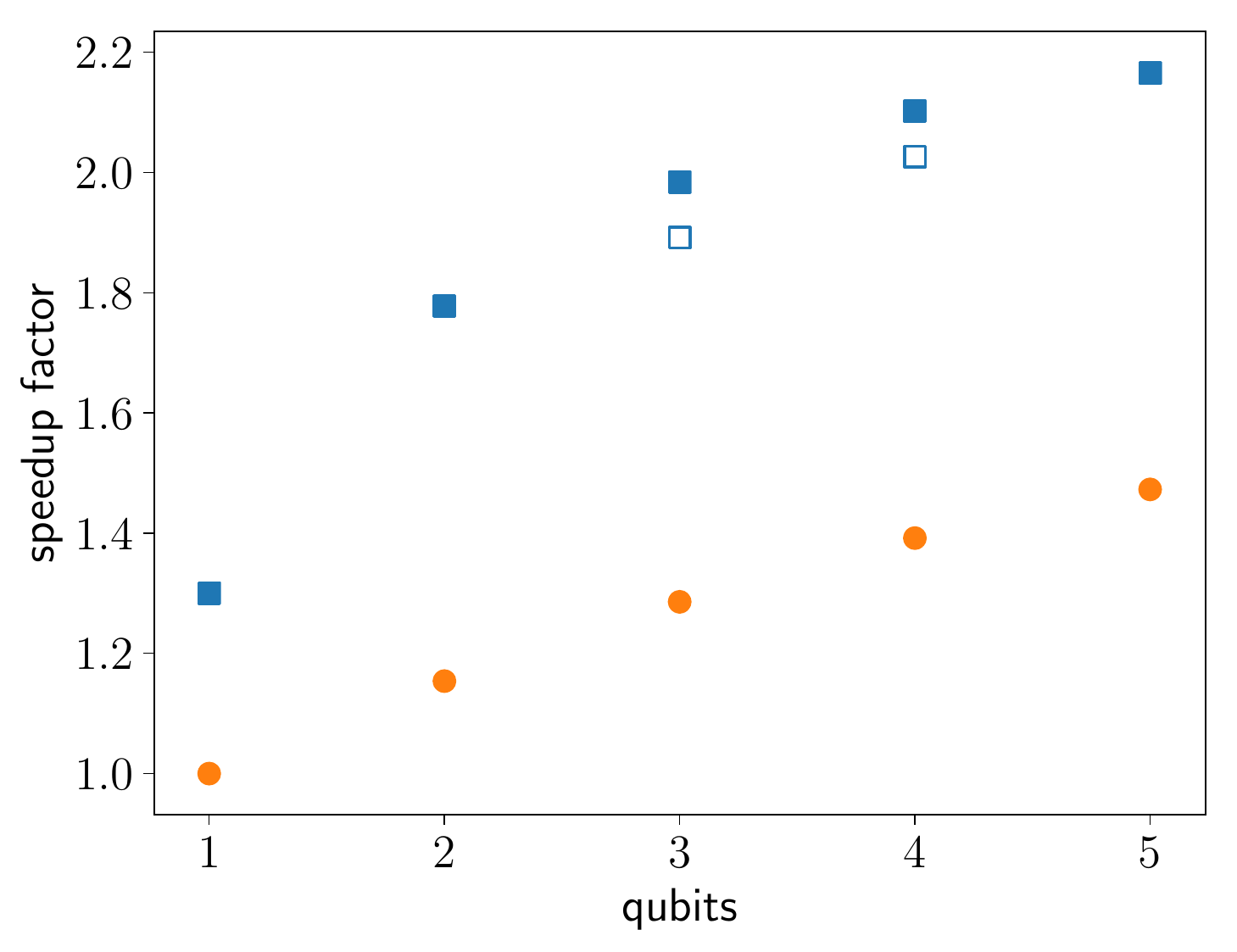}
}
\caption{The speedup factor. The six-state scheme (solid blue square) achieves greater
  values of speedup than the three-base scheme (solid orange circle). Empty blue
  squares represent the values measured in the three-qubit state tomography and
  the two-qubit process characterization, respectively.}
\label{fig:speedup}
\end{figure}

The total angular (and thus, temporal) duration of tomographic tasks increases exponentially with the size of the quantum system. However, the TSP optimization of the order of measurements yields reduction of the total tomography duration, shown in Fig.~\ref{fig:tau}, which also scales exponentially. The only exception is the one-qubit, three-base tomography, where the conventional order is optimal already. The speedup factor, plotted in Fig.~\ref{fig:speedup}, is generally bigger for the six-state scheme compared to the three-base scheme, but in both cases it follows the same monotonously rising trend, reaching a value of 2 already for three-qubit tomography relying on the six-state scheme.
The limitation on the number of qubits for which we were able to find the optimal
measurement orders seems to originate from the size of the problem, namely the
number of the measurements required for complete tomography. In the six-qubit case,
this is 46,656, which is out of reach for the state-of-the-art TSP solvers.
Additional concerns come from the memory limitations of the consumer-grade computers
we used the TSP solver on. However, even for larger systems, where the complete optimization cannot
be performed, it is still possible to utilize the results for a smaller
subsystem and achieve the corresponding speedup. For example, six-qubit
tomography can be partially optimized by setting the measurements on
the five-qubit subsystem in the optimal order and changing the sixth qubit
measurement in the conventional order. This approach would still yield
the speedup factor corresponding to a five-qubit system.

To demonstrate the speedup under real conditions, 
we have used the TSP solver to optimize the order of preparations and measurements in
existing experimental setups. We performed quantum state tomography of a three-qubit
entangled Greenberger--Horne--Zeilinger state prepared by a Toffoli gate
(controlled-controlled-NOT gate) \cite{Micuda2013}.
Furthermore, quantum process tomography was used to fully characterize
a two-qubit SWAP gate \cite{Starek2018}.
The computer program responsible for manipulating the rotation mounts was modified
to record timestamps just before and after each transition. We were
thus able to compute the speedup factor, and arrived at results close to the
predicted values (see Fig.~\ref{fig:speedup}).
The slight discrepancy between the predicted and the measured values of the speedup
is caused by time overhead of the communication with the motorized rotation mounts.
Also, small deviation from linear dependence of mounts' rotation time and the angle
travelled contributes to this discrepancy.

An important fact to note is that we have compared the total times spent on
reorienting the wave plates, not the total time of tomography. When comparing
the latter, we may arrive at a smaller speedup factor. This is due to the fact that
the tomography consists of other operations, apart from the wave plate readjustment,
such as data acquisition and hardware communication overhead.
Using the optimal order of measurements and a low-latency multi-channel counter
\cite{Countex2018}, we have been able to reduce the overall time of full quantum
process tomography of a three-qubit quantum controlled-controlled-phase gate
\cite{Starek2018b} from approximately 23 hours to less than 11 hours.
Consequently, the impact of setup instabilities has been effectively reduced.

\section{Different tomography schemes and information encodings}

\begin{figure}[t!]
\centering{\includegraphics[width=0.7\columnwidth]{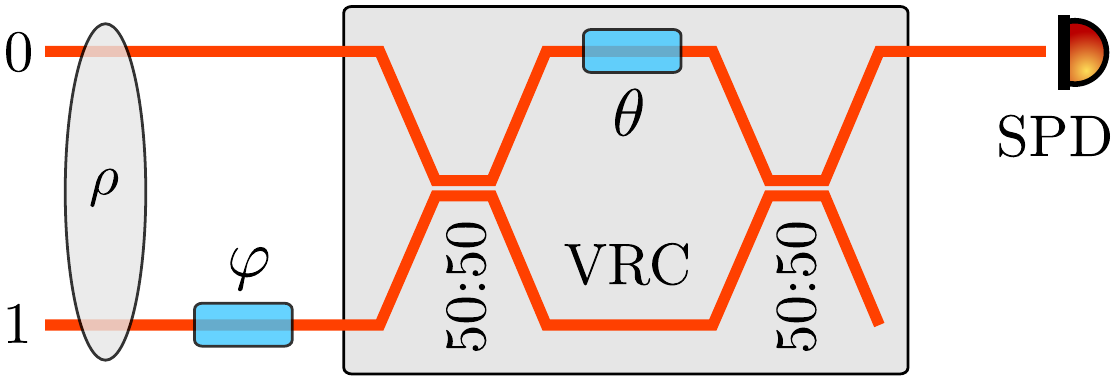}}
\caption{The scheme of on-chip tomography characterization of path-encoded qubit in quantum state $\rho$. The phase $\varphi$ of qubit projection measurement is set by an heating element. The second heater affects the phase $\theta$ in the Mach-Zehnder interferometer acting as a variable ratio coupler (VRC).}
\label{fig:pathenc}
\end{figure}

\begin{figure*}[t!]
\centering{
{\includegraphics[width=0.7\textwidth]{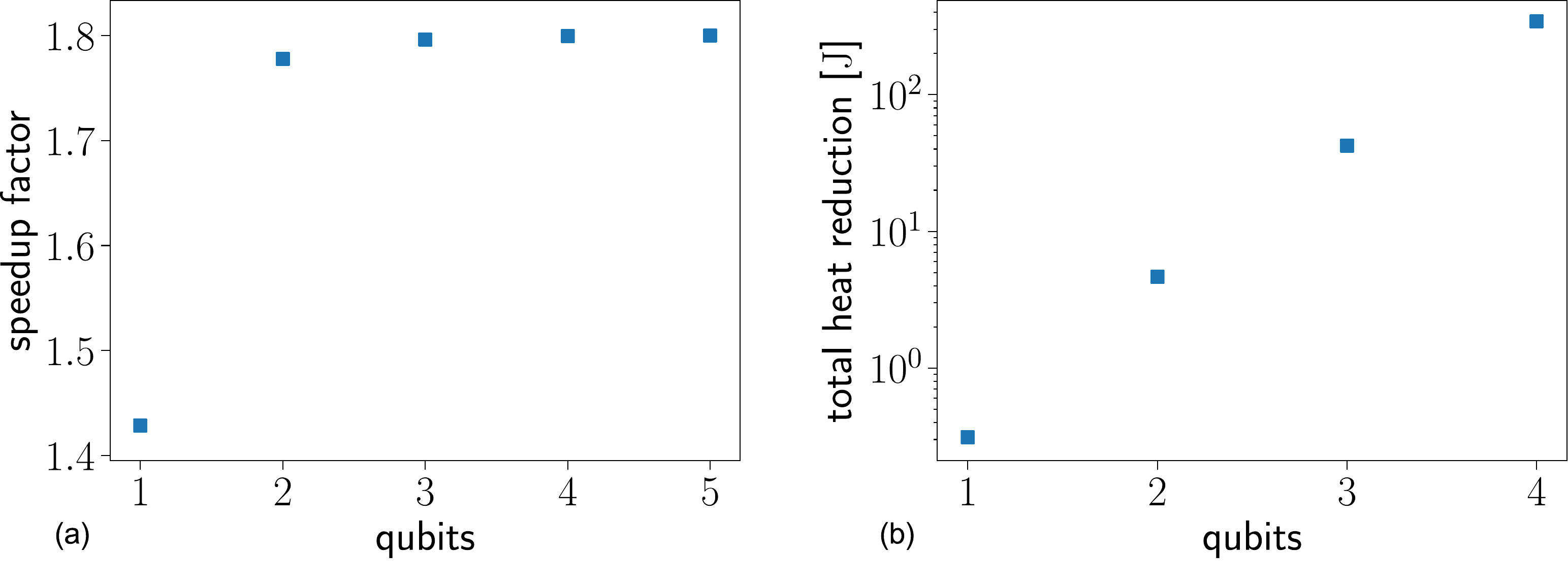}
\label{fig:fig5b}}
}
\caption{
The speedup factor (a) for temporal optimization of the path-encoded state tomography using heater elements practically saturates at the value of 1.8 for more than two qubits. The total heat reduction (b) for heat-optimized tomography, assuming that power of $0.5$ W applied over 1 second is required for a $2\pi$ phase change.
}
\label{fig:fig5}
\end{figure*}

To show the wide applicability of the optimization strategy we also demonstrate tomography speedup for path-encoded quantum circuitry on an optical chip. The path-encoded qubit is formed by two paths of a Mach-Zehnder interferometer with different probabilities of having a photon in the upper (0) and lower (1) arms, and a relative phase between them. The key element here is a phase shifter, often implemented using a resistive heating element on the chip surface, enabling reconfigurability of the waveguide circuit \cite{OBrien2009,Walmsley2009,Osellame2015,Laing2015,Sciarrino2017}.
The preparation of input states and the projection measurements requires setting the phase and changing the splitting ratio of a waveguide coupler. The former can be performed with the help of the heating element but the latter cannot be achieved easily on the chip. Instead, a Mach-Zehnder interferometer with the heater in one arm is employed to emulate the functionality of a variable-ratio coupler, see Fig.~\ref{fig:pathenc}. The required phase settings for six-state tomography with a single detector per qubit are shown in Table \ref{tab:pathenc}. The phase induced by the heater, and also the required settling time are proportional to the dissipated power, with approximately $0.5$ W inflicting the phase change of $2\pi$ within 1 s \cite{Osellame2015}.
Two different optimization goals can be pursued, leading to minimization of either the total time of phase readjusting or the total heat transferred to the chip from all heaters. The time minimization problem is formulated in phase units (rad) in a similar way as for polarization tomography, with the adjacency matrix elements given by the maximum phase change across all the involved heaters. The achievable speedup reaches 1.43 for 1 qubit and practically saturates at the value of 1.80 for two and more qubits, see Fig.~\ref{fig:fig5}(a).
The thermal load minimization problem depends severely on the exact temporal response of the heater. For simplicity we assume that the temperature (and the inflicted phase change) increases linearly with time after the heating power is switched on and before the target temperature (phase) is reached. Consequently, the total heat transferred is proportional to the product of the power being dissipated by all the heaters and the settling time of the heater that is responsible for the largest phase change. Even such simplified description requires the use of nontrivial asymmetric adjacency matrices. The optimization yields heat reduction of 0.3 J (reduction factor of 1.59) for a single qubit, which goes up up to 343 J (reduction factor of 1.79) for four qubits, see Fig.~\ref{fig:fig5}(b). Upon comparing the total duration of the heat-optimized measurement sequences with the time-optimized ones, we found that they were not significantly different. This leads us to the conclusion that both duration and heat reduction can be achieved at once.

\begin{table}[t!]
	\centering
	\caption{One-qubit path-encoded tomography (the six-state scheme). The 0, 1,
    $+$, $-$, i, and $-$i projections are shown with the corresponding phase settings.}
    \medskip
\begin{tabular}{c|ccc}
~                  & T:R   & $\theta$ & $\varphi$ \\ \hline
$| 0\rangle$           & 100:0 & 0        & 0         \\
$| 1\rangle$           & 0:100 & $\pi$    & 0         \\
$| +\rangle$           & 50:50 & $\pi/2$  & 0         \\
$| -\rangle$           & 50:50 & $-\pi/2$ & 0         \\
$| \textrm{i}\rangle$  & 50:50 & $\pi/2$  & $\pi/2$   \\
$| -\textrm{i}\rangle$ & 50:50 & $\pi/2$  & $-\pi/2$ 
\end{tabular}
    \label{tab:pathenc}
\end{table}

The tomography schemes studied so far consist of probing by and projecting to tensor products of the eigenstates of Pauli operators. However, square-root measurements \cite{Wootters1996}, random sampling \cite{Flammia2011}, and compressed sensing techniques \cite{Eisert2010,White2011,Eisert2017} benefit from random (or at least randomly selected) projection measurements. The optimization of all the schemes in the plethora of measurement frameworks utilizing some form of randomness in state preparation or measurement is beyond the scope of this work. To simply demonstrate the applicability of the proposed TSP optimization strategy, we show the average speedup reached by the optimal reordering of polarization projection measurements with randomly generated rotation angles of the half-wave and quarter-wave plates. The total number of individual measurements were chosen to be the same as for the previous tomographic scheme based on tensor products of single-qubit Pauli eigenstates to facilitate the comparison with the previous results. The random-measurement scheme speedup, averaged from ten different sets of randomly selected wave plate angles, reaches 2.18 for two qubits already.

\section{Conclusion}

We have shown that the total duration of quantum tomography can be considerably
reduced by an optimal reordering the constituent measurements.
We have presented the solutions and experimental verifications
of the devised optimization strategy for photonic quantum tomography
of up to six polarization encoded qubits.
The total time spent on reorienting the polarization-changing elements
required for the tomography procedure is minimized, which yields nearly
twofold increase in speed already for three-qubit state tomography.
The speedup has been analyzed also for advanced input-output process
characterization with the optimized sequence of probe state preparations and
output measurements. As an example, the duration of full tomographic
characterization of a complex quantum logic gate, consisting of 47 thousands
constituent measurements, was reduced to less than 50\%, compared
to the unoptimized conventional strategy.

The optimization strategy presented here for polarization encoding with
wave plates can also be used with other analyzers, encodings, and even
different optimization goals in mind. For example, on-chip path-encoding
utilizing heating elements to change the phase between various paths
can be optimized not only to decrease the total measurement duration
but also to diminish the overall thermal load.
Furthermore, the reported tomography optimization is applicable to
continuous-variable \cite{SanchezSoto2012} or high-dimensional
systems \cite{Sciarrino2016}.
As with the full quantum tomography, also other measurement schemes like
direct fidelity estimation \cite{Flammia2011}, permutationally invariant
tomography \cite{Weinfurter2010}, and matrix-product-state tomography \cite{Poulin2010,Guo2017}
can benefit from the presented optimization approach.
The achievable speedup increases with the number of parties and
preparation/measurement settings.

\section*{Funding}
Czech Science Foundation (project 17-26143S);
MEYS and European Union's Horizon 2020 (2014-2020) research and innovation framework programme under grant agreement No 731473 (project HYPER-U-P-S, No 8C18002);
Palack\'y University (project IGA-PrF-2018-010).




%

\end{document}